\documentclass[letterpaper,prl,showpacs,twocolumn]{revtex4}
\usepackage{bm}
\usepackage{graphicx}
\begin{document}

\title{Formation kinetics of magnetic chains, rings, X's, and Y's}

\author{Peter D. Duncan}

\author{Philip J. Camp}

\email{philip.camp@ed.ac.uk}

\affiliation{School of Chemistry, University of Edinburgh, West Mains
Road, Edinburgh EH9 3JJ, United Kingdom}

\date{\today}


\begin{abstract} The kinetics of aggregation in a monolayer of magnetic
particles are studied using stochastic dynamics computer simulations. At
low densities ($\leq 8\%$ coverage) the equilibrium structure is made up
of chains and rings; the primary mechanisms by which these motifs form are
described. At higher densities ($> 15\%$ coverage), we observe large
transient concentrations of Y-shaped and X-shaped defects that ultimately
give way to an extended, labyrinthine network. Our results suggest that a
defect mechanism -- such as that proposed by Tlusty and Safran, {\em
Science} {\bf 290}, 1328 (2000) -- could drive a metastable phase
separation in two dimensions. \end{abstract}

\maketitle

The structure, phase behavior, and dynamics of strongly interacting,
dipolar fluids present considerable challenges to soft-matter physics. The
most common realization of a dipolar fluid is a ferromagnetic colloidal
suspension, or ferrofluid. In the ideal case, this consists of spherical,
homogeneously magnetized monodisperse particles with diameters $\sim
10~\mbox{nm}$, sterically stabilized and immersed in a nonpolar solvent.
The resulting colloidal interactions are caricatured by the widely studied
dipolar hard sphere fluid. Despite almost four decades of intensive
experimental, theoretical, and simulation study
\cite{Teixeira:2000/a,Holm:2005/a}, at least one outstanding question
remains to be answered definitively: are point dipolar interactions alone
sufficient to drive vapor-liquid phase separation?

On the one hand, the Boltzmann-weighted, angle average of the
dipole-dipole potential gives (to leading order) an isotropic, attractive
pair potential that varies like $-1/r^{6}$, where $r$ is the interparticle
separation; this is expected to produce conventional condensation behavior
\cite{deGennes:1970/a}. On the other hand, simulations show that
conventional condensation is preempted by strong aggregation, driven at
low temperatures by the energetically favorable `nose-to-tail'
conformation \cite{Weis:1993/a}. If phase separation occurs in 3D, then it
is of a rather unusual nature; simulations suggest that the low-density
phase mainly consists of chain-like aggregates, while the high-density
phase resembles a network of interconnected segments \cite{PJC:2000/a}.  
One possible scenario involves a defect-mediated phase transition
\cite{Tlusty:2000/a} in which the chains' defects are the singly-connected
particles at the chain ends, while the network's defects consist of
particles with three or four near neighbors in a Y-shaped or X-shaped
conformation, respectively. Such fundamental issues are not only of
relevance to magnetic fluids; the physical properties of many materials
are governed by connectivity and topology \cite{Zilman:2003/a}.

Recently, 2D dipolar fluids (with 3D magnetostatics) have received
attention due to the possibility of directly imaging aggregation in thin
films using cryogenic transmission electron microscopy
\cite{Puntes:2001/a,Butter:2003/a,Klokkenburg:2006/a}. The equilibrium
structure \cite{Tavares:2002/a,Weis:2003/a} and dynamics \cite{PJC:2004/d}
of 2D dipolar fluids have been studied in detail using computer
simulations. At low density and low temperature, the dominant structural
motifs are isolated chains and rings. The high-density structure consists
of a labyrinthine network of long chains, with a small concentration of
X-shaped and Y-shaped defects. There has never been any suggestion of a
vapor-liquid phase transition in the 2D system, but a transition between
isolated and system-spanning clusters at low-density has recently been
characterized \cite{Tavares:2006/a}. The relative ease with which complex
self-assembled structural motifs can be visualized and analyzed in
experiments and simulations means that 2D fluids are attractive and
important objects of study. Given the significance of such systems, it is
surprising that more isn't known about the self-assembly process itself,
starting from a `random' arrangement of particles. In one of the few
studies in this area, Wen {\it et al.} directly imaged the aggregation of
nickel-plated glass microspheres \cite{Wen:1999/a}. They observed that
rings can be formed by two short chains joining at both ends
simultaneously. Branching at X and Y defects was not considered.

In this Letter, we present a detailed simulation study of the aggregation
process in monolayers of strongly dipolar particles. Starting from
equilibrated configurations of non-polar particles, we elucidate the
mechanisms of cluster formation that occur when the dipoles are `switched
on'. At low densities, rings are predominantly formed by single, isolated
chains folding up (rather than by the association of two short chains).
Interestingly, at high densities, the system shows a high transient
concentration of defect particles. This is significant because it suggests
that highly branched structures could be stabilized kinetically. It might
therefore be possible to realize a metastable, defect-mediated phase
separation in the laboratory. We also briefly consider the types of
mechanisms by which the structure evolves at long times.

We model the system as a monolayer of monodisperse dipolar soft spheres.  
The interparticle potential is given by
\begin{equation}
u(r,\bm{\mu}_{1},\bm{\mu}_{2}) = 
4\epsilon\left(\frac{\sigma}{r}\right)^{12} +
 \frac{\bm{\mu}_{1}\cdot\bm{\mu}_{2}}{r^{3}}
-\frac{3(\bm{\mu}_{1}\cdot\bm{r})(\bm{\mu}_{2}\cdot\bm{r})}{r^{5}}
\label{eqn:pot}
\end{equation}
where $\epsilon$ is an energy parameter, $\sigma$ is the sphere diameter,
$\bm{\mu}_{i}$ is the dipole vector on particle $i$, $\bm{r}$ is the
interparticle separation vector, and $r=|\bm{r}|$. Reduced units are
defined as follows: temperature $T^{*}=k_{\rm B}T/\epsilon$, where $k_{\rm
B}$ is Boltzmann's constant; dipole moment
$\mu^{*}=\sqrt{\mu^{2}/\epsilon\sigma^{3}}$; particle number density
$\rho^{*}=\rho\sigma^{2}$, where $\rho=N/L^{2}$ and $L$ is the length of
the square simulation cell; time $t^{*}=t/\tau$ where
$\tau=\sqrt{m\sigma^{2}/\epsilon}$ is the basic unit of time. Stochastic
dynamics simulations were performed according to the integrated Langevin
equations \cite{Jones:1992/a}
\begin{eqnarray}
\bm{r}_{i}(t+\delta t) &=& \bm{r}_{i}(t) 
+ \frac{D_{0}^{t}}{k_{\rm B}T} 
  \bm{F}_{i} \delta t
+ \delta\bm{W}_{i}^{t}
\label{eqn:eomr} \\
\hat{\bm{\mu}}_{i}(t+\delta t) &=& \hat{\bm{\mu}}_{i}(t)
+ \frac{D_{0}^{r}}{k_{\rm B}T} 
  \bm{T}_{i} \wedge \hat{\bm{\mu}}_{i}(t) \delta t
+ \delta\bm{W}_{i}^{r} \wedge \hat{\bm{\mu}}_{i}(t) \nonumber \\
\label{eqn:eommu}
\end{eqnarray}
where $\bm{F}_{i}$ ($\bm{T}_{i}$) is the net force (torque) acting on
dipole $i$ at time $t$, $\hat{\bm{\mu}}_{i}=\bm{\mu}_{i}/\mu$ is a unit
dipole orientation vector, $\delta t$ is the integration time step, and
$D_{0}^{t}$ ($D_{0}^{r}$) is the translational (rotational) diffusion
constant at infinite dilution. The components of the 2D vector
$\delta\bm{W}_{i}^{t}$ and the 3D vector $\delta\bm{W}_{i}^{r}$ were
generated independently from Gaussian distributions subject to the
conditions $\langle\delta \bm{W}_{i}^{t}\rangle = 0$, $\langle \delta
\bm{W}_{i}^{t} \cdot \delta \bm{W}_{j}^{t} \rangle = 4D_{0}^{t}\delta t
\delta_{ij}$, $\langle\delta \bm{W}_{i}^{r}\rangle = 0$, and $\langle
\delta \bm{W}_{i}^{r} \cdot \delta \bm{W}_{j}^{r} \rangle = 6D_{0}^{r}
\delta t \delta_{ij}$. In this scheme, the short-time inertial dynamics
are suppressed: with high dipole moments these occur on timescales of
order $1$ (in reduced units) \cite{PJC:2004/d}, while of primary interest
here are conformational processes occurring on timescales orders of
magnitude longer.

We present results for $N=1024$ particles at a range of densities, with a
large dipole moment $\mu^{*}=2.75$, and temperature $T^{*}=1$.
Characteristic diffusion constants were estimated from the (stick)
Stokes-Einstein laws yielding $D_{0}^{t} = k_{\rm B}T/3\pi\eta \sigma
\simeq 4 \times 10^{-11}~\mbox{m}^{2}~\mbox{s}^{-1}$ and $D_{0}^{r} =
k_{\rm B}T/\pi\eta \sigma \simeq 1 \times 10^{6}~\mbox{s}^{-1}$ for
spherical particles with $\sigma=10~\mbox{nm}$, in a solvent of viscosity
$\eta=10^{-3}~\mbox{Pa}~\mbox{s}$ at temperature $T=300~\mbox{K}$; the
dimensionless quantities $D_{0}^{t}\tau/\sigma^{2}=0.004$ and
$D_{0}^{r}\tau=0.01$ were obtained using the mass for $10~\mbox{nm}$
spheres with mass density $\sim 8000~\mbox{kg}~\mbox{m}^{-3}$ (typical for
iron or cobalt) and energy parameter $\epsilon=k_{\rm B}T$. The
integration time step was $\delta t^{*}=0.01$. Self-assembly was initiated
from configurations generated with $\mu^{*}=0$. For each density studied,
five independent runs with different initial configurations were
conducted, and the results for each density were averaged. The
configurational temperatures associated with the positions and
orientations of the particles were measured independently
\cite{Chialvo:2001/a}. In all cases the instantaneous configurational
temperatures fluctuated about $T^{*}=1$, with rms deviations of about
$0.1$, throughout the self-assembly process.

\begin{figure}[tb] \includegraphics[width=8.6cm]{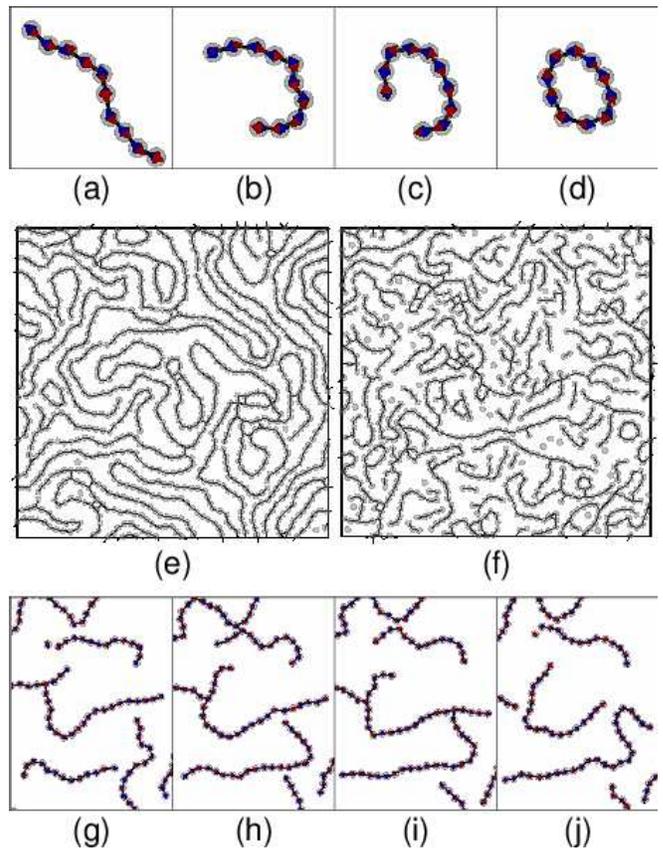}
\caption{\label{fig:snapshots} (Color online) Instantaneous configurations
at various densities and times:  (a) $\rho^{*}=0.05$, $t^{*}=3500$;  (b)
$\rho^{*}=0.05$, $t^{*}=4330$;  (c) $\rho^{*}=0.05$, $t^{*}=4800$;  (d)  
$\rho^{*}=0.05$, $t^{*}=4950$; (e) $\rho^{*}=0.5$, $t^{*}=10000$;  (f)
$\rho^{*}=0.5$, $t^{*}=50$; (g) $\rho^{*}=0.2$, $t^{*}=1080$;  (h)
$\rho^{*}=0.2$, $t^{*}=1200$;  (i) $\rho^{*}=0.2$, $t^{*}=1300$;  (j)
$\rho^{*}=0.2$, $t^{*}=1400$. The black lines between particles denote
`bonds' identified using an energy criterion. In (a)-(d) and (g)-(j) the
orientations of the dipole moments are indicated with red and blue cones.}
\end{figure}

At equilibrium, the low-density structure mainly consists of small rings
and chains \cite{PJC:2004/d}. In Figs.~\ref{fig:snapshots}(a)-(d) we show
how a ring is formed by a chain closing in on itself, at a density
$\rho^{*}=0.05$. Wen {\it et al.} suggest that rings are formed by two
short chain-like segments making connections at either end simultaneously
\cite{Wen:1999/a}. In contrast, movies of the aggregation process in our
simulations show that the majority of rings are formed from isolated
chains. At high density the equilibrium structure resembles a labyrinth of
long, winding chains \cite{PJC:2004/d}; an `equilibrium' configuration at
$\rho^{*}=0.5$ is shown in Fig.~\ref{fig:snapshots}(e). Locally, the
equilibrium structures at low and high density are not so different; most
particles are in chain-like environments, flanked by two near neighbors.

The process of self-assembly was monitored by identifying particles
belonging to the same cluster on the basis of an energy criterion. We
chose an energy cut-off of $u_{c}=-0.6{\mu^{*}}^{2}$, which captures a
range of likely conformations for neighboring dipoles, either in chains or
in more exotic defect environments \footnote{Earlier work on the
equilibrium clusters in dipolar hard sphere fluids employed a distance
criterion \cite{Tavares:2002/a}. The optimum cut-off depends sensitively
on the short-range potential.  Nonetheless, we have confirmed that a
distance cut-off of $1.5\sigma$ yields comparable results to those
reported here.}. We identified `terminal particles' (one near neighbor),
`internal particles' (two near neighbors), `defect particles' (three or
more near neighbors), chains, rings, and `defect clusters' (containing at
least one defect particle).

\begin{figure}[tb] \includegraphics[width=8.6cm]{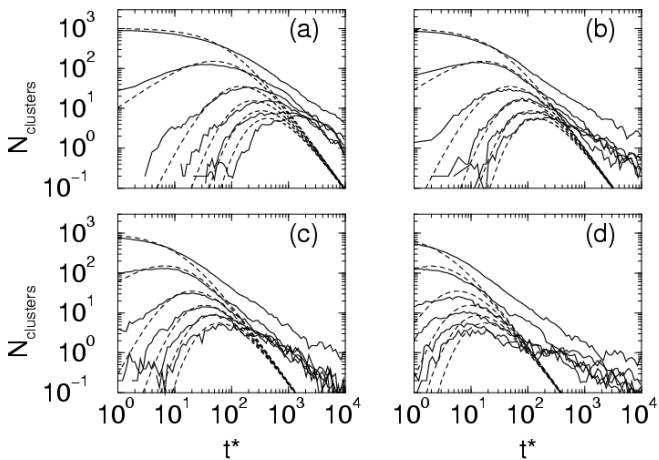}
\caption{\label{fig:clusters} Numbers of clusters containing (from top to
bottom) $n=1$, $2$, $4$, $6$, $8$, or $10$ particles, as functions of
time: (a) $\rho^{*}=0.1$; (b) $\rho^{*}=0.2$; (c) $\rho^{*}=0.3$; (d)
$\rho^{*}=0.5$. The solid lines are simulation results, and the dashed
lines are fits to the von Smoluchowski model.} \end{figure}

In Fig.~\ref{fig:clusters} we plot the numbers of $n$-mers ($n=1$, $2$,
$4$, $6$, $8$, and $10$) as functions of time at densities in the range
$0.1 \leq \rho^{*} \leq 0.5$ \footnote{We also studied $\rho^{*}=0.05$,
but the results are very similar to those at $\rho^{*}=0.1$ and are
therefore omitted for brevity.}. To analyze the simulation results, we use
a simple von Smoluchowski model \cite{Sonntag:1987/a}
\begin{equation}
\frac{{\rm d}\rho_{n}}{{\rm d}t} =
\frac{1}{2}\sum_{i+j=n}k_{ij}\rho_{i}\rho_{j} -
\rho_{n}\sum_{i=1}^{\infty}k_{ni}\rho_{i} 
\label{eqn:smol} 
\end{equation}
where $\rho_{n}$ is the number density of $n$-mers, and the $k_{ij}$'s are
`rate constants'. Under the simplifying assumption that all rate constants
are equal ($k_{ij}=k$), the number of $n$-mers is $N_{n} = N (k\rho
t)^{n-1} / (1+k\rho t)^{n+1}$, and the total number of clusters is
$\sum_{n=1}^{\infty}N_{n} = N/(1+k\rho t)$, where $\rho$ is the total
number density of particles. We determined a reduced rate constant
$k^{*}=k\tau/\sigma^{2}$ for each density by fitting to the total number
of clusters, with the results $k^{*}=0.049(3)$, $0.104(5)$, $0.156(4)$,
$0.247(6)$, and $0.52(1)$ at $\rho^{*}=0.05$, $0.1$, $0.2$, $0.3$, and
$0.5$, respectively. The resulting curves for $N_{n}(t)$ are shown in
Fig.~\ref{fig:clusters}. The general level of agreement is quite good at
short times, but at longer times the von Smoluchowski model underestimates
the numbers of clusters. This is likely to be due to complex and slow
relaxational processes occurring as the structure ripens, and the gross
simplification of setting all rate constants equal.

\begin{figure}[tb] \includegraphics[width=8.6cm]{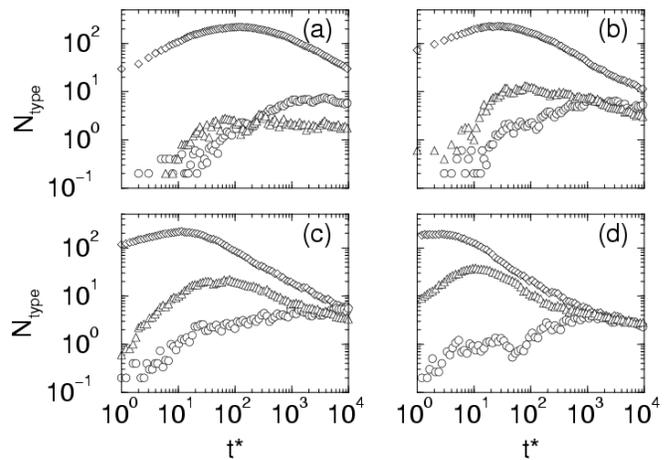}
\caption{\label{fig:type} Numbers of chains (excluding monomers)
(diamonds), rings (circles) and defect clusters (triangles), as functions
of time, at densities (a) $\rho^{*}=0.1$, (b) $\rho^{*}=0.2$, (c)
$\rho^{*}=0.3$, and (d) $\rho^{*}=0.5$.} \end{figure}

In Fig.~\ref{fig:type} we show the numbers of chains, rings, and defect
clusters as functions of time at densities in the range $0.1 \leq \rho^{*}
\leq 0.5$. At all densities, the numbers of chains are greater than the
numbers of rings and defects during the early stages of the aggregation
process, but the long-time behaviors are very different. Rings are favored
over defect clusters at low density ($\rho^{*} \leq 0.1$), in accord with
the known equilibrium structures. At high density ($\rho^{*}\geq 0.3$)
defect clusters are favored over rings; the numbers of defect clusters
show strong maxima at intermediate times ($t^{*} \sim 10$-$100$). At
intermediate density ($\rho^{*}=0.2$) there are roughly equal numbers of
rings and defect clusters.

\begin{figure}[tb] \includegraphics[width=8.6cm]{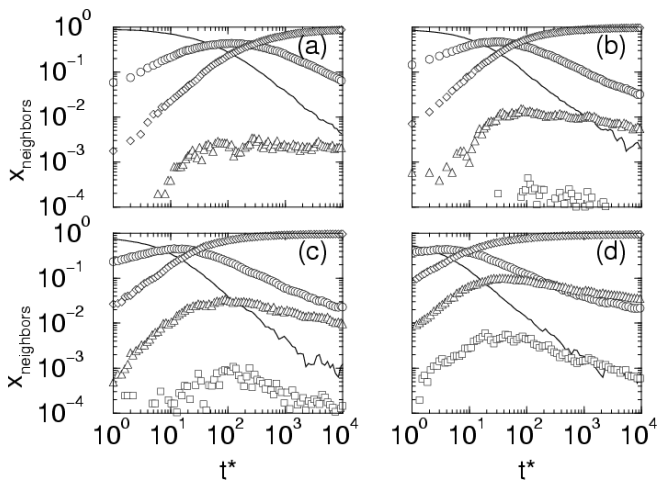}
\caption{\label{fig:neighbors} Fractions of particles with $n$ neighbors,
as functions of time, at densities (a) $\rho^{*}=0.1$, (b) $\rho^{*}=0.2$,
(c) $\rho^{*}=0.3$, and (d) $\rho^{*}=0.5$: $n=0$ (solid line); $n=1$
(circles); $n=2$ (diamonds); $n=3$ (triangles); $n=4$ (squares).}
\end{figure}

To further elucidate the aggregation mechanisms, in
Fig.~\ref{fig:neighbors} we plot $x_{n}(t)$, the fraction of particles
with $n$ neighbors ($n=0$-$4$) at time $t$, at densities in the range $0.1
\leq \rho^{*} \leq 0.5$. (No particles with five or more near neighbors
were observed.) At all densities, $x_{0}$ falls monotonically, $x_{1}$
shows a maximum, and $x_{2} \rightarrow 1$ as the majority of particles
ultimately end up with two neighbors as parts of chains or rings. Dramatic
differences between low-density and high-density aggregation kinetics are
evidenced by $x_{3}$ and $x_{4}$. At low density ($\rho^{*} \leq 0.1$)
less than $\sim 0.2\%$ of particles have three or more neighbors at any
given time. At intermediate density ($\rho^{*}=0.2$)  $x_{3}$ shows a
maximum at $t^{*} \sim 100$, while $x_{4}$ is essentially negligible. At
higher densities ($\rho^{*}=0.3$, $0.5$) and at intermediate times
($t^{*}\sim 100$), up to $10\%$ of all particles are defects, with
approximately ten times more particles having three neighbors than four
neighbors. In Fig.~\ref{fig:snapshots}(f) we show a snapshot of a
configuration at $\rho^{*}=0.5$ and $t^{*}=50$, i.e., close to the
location of the peak in $x_{3}$. Note the higher connectivities of
particles within clusters as compared to those when the system is closer
to equilibrium [$t^{*}=10000$, Fig.~\ref{fig:snapshots}(e)]. Ultimately,
the number of defect particles falls again by at least an order of
magnitude as equilibrium is approached.

Of course, aggregates continue to disassemble and form, even at
equilibrium. During the aggregation process at high density, there is a
very slow net decrease in the number of defect particles within the
system. It is impossible to identify specific local events where the
number of defects is reduced irreversibly, but in
Figs.~\ref{fig:snapshots}(g)-(j) we show a sequence of snapshots that
illustrate the kinds of processes which, in the long run, lead to a
changing cluster distribution. The time interval $t^{*}=1080$-$1400$
corresponds to where $x_{3}$ is decreasing relatively rapidly. In the
upper regions of these figures, we see how two chains segments meet, bond
temporarily, and then separate. In the lower halves of
Figs.~\ref{fig:snapshots}(g) and \ref{fig:snapshots}(h), we see how two
chains can collide at right angles, and split in to two new chains. In
Fig.~\ref{fig:snapshots}(i), we see a cluster with two defect particles;  
in Fig.~\ref{fig:snapshots}(j) we see the result of that cluster having
ruptured at the positions of those two defects almost simultaneously.

In summary, our results show that the aggregation mechanisms at low
density and high density are very different. At low density, aggregation
proceeds through the formation of chains, some of which will go on to form
rings, as shown in Fig.~\ref{fig:snapshots}(a)-(d). At high density, it
appears that the transient aggregates include a significant proportion of
defect clusters, such as those implicated in 3D dipolar phase separation
\cite{Tlusty:2000/a}. At equilibrium, the particle connectivities at low
density and high density are somewhat similar in that the vast majority of
particles have exactly two neighbors.

Our observations suggest a direct experimental test for the theory of
defect-mediated phase transitions put forward by Tlusty and Safran
\cite{Tlusty:2000/a}. If the ferrocolloid particles were modified to
introduce a strong, short-range attraction (e.g. chemically, or with added
polymer to induce depletion forces) then the transient network structure
at high density may be kinetically stabilized long enough for phase
separation to occur, even in thin films where no equilibrium transition is
anticipated. This can only arise if the high-density branched structure
corresponds to a {\em local} free-energy minimum. The experimental
approach suggested here could provide a way of guiding the dense phase in
to that minimum, but there may be alternative strategies. In any case, we
hope that aggregation kinetics and metastable phases in strongly dipolar
fluids can be investigated by direct experimental observation.

We thank the School of Chemistry at the University of Edinburgh for the
provision of an EPSRC DTA studentship to PDD.
 

\end{document}